\documentstyle[12pt]{article}
\newlength{\pubnumber} 

\catcode`\@=11
\@addtoreset{equation}{section}

\def\section{\@startsection{section}{1}{\z@}{3.5ex plus 1ex minus .2ex}
 {2.3ex plus .2ex}{\large\bf}}
\def\subsection{\@startsection{subsection}{2}{\z@}{2.3ex plus .2ex}
 {2.3ex plus .2ex}{\bf}}

\begin{document}

\begin{titlepage}
\samepage{
\rightline{CERN-TH/99-267}
\rightline{\tt hep-ph/9908530}
\rightline{August 1999}
\vfill
\begin{center}
  {\Large \bf Anomaly-Induced Gauge Unification 
             and Brane/Bulk Couplings
              in Gravity-Localized Theories\\}

\vfill
   {\large
    Keith R. Dienes$^{1,2}$\footnote{
     E-mail address: keith.dienes@cern.ch},
    Emilian Dudas$^{3}$\footnote{
     E-mail address:  emilian.dudas@th.u-psud.fr},
     $\,$and$\,$
      Tony Gherghetta$^1$\footnote{
     E-mail address: tony.gherghetta@cern.ch}
    \\}
\vspace{.18in}
 {\it  $^1$ Theory Division, CERN, CH-1211 Geneva 23, Switzerland\\}
\vspace{.04in}
 {\it  $^2$ Department of Physics, University of Arizona, 
             Tucson, AZ  85721 USA\\}
\vspace{.04in}
 {\it  $^3$ LPTHE, Univ.\ Paris-Sud, F-91405, Orsay Cedex, France\\}
\end{center}
\vfill
\begin{abstract}
  {\rm
      It has recently been proposed that gravity-localized
      compactifications can generate the required gauge
      hierarchy without the need for hierarchically large 
      extra spacetime dimensions.
      In this paper, we show that
      gauge coupling unification arises naturally in 
      such scenarios as a result of the anomaly induced by 
      the rescaling of the wavefunctions of the brane fields.  
      Thus, ``anomaly-induced'' gauge coupling unification
      can easily explain the apparent low-energy gauge couplings 
      in gravity-localized compactifications.
      However, we also point out a number of phenomenological 
      difficulties with such compactifications, including an 
      inability to accommodate the GUT scale and the electroweak 
      scale simultaneously.  We also show that brane/bulk couplings
      in this scenario are generically too small to be
      phenomenologically relevant.  
      Finally, we speculate on possible resolutions to these puzzles.
          } 
\end{abstract}
\vfill
\smallskip}
\end{titlepage}

\setcounter{footnote}{0}

\newcommand{\newc}{\newcommand}

\newc{\gsim}{\lower.7ex\hbox{$\;\stackrel{\textstyle>}{\sim}\;$}}
\newc{\lsim}{\lower.7ex\hbox{$\;\stackrel{\textstyle<}{\sim}\;$}}

\def\beq{\begin{equation}}
\def\eeq{\end{equation}}
\def\beqn{\begin{eqnarray}}
\def\eeqn{\end{eqnarray}}
\def\sosixteen{{$SO(16)\times SO(16)$}}
\def\e8{{$E_8\times E_8$}}
\def\V#1{{\bf V_{#1}}}
\def\half{{\textstyle{1\over 2}}}
\def\ttwo{{\vartheta_2}}
\def\tthree{{\vartheta_3}}
\def\tfour{{\vartheta_4}}
\def\ttwob{{\overline{\vartheta}_2}}
\def\tthreeb{{\overline{\vartheta}_3}}
\def\tfourb{{\overline{\vartheta}_4}}
\def\etainv{{\overline{\eta}}}
\def\Str{{{\rm Str}\,}}
\def\bone{{\bf 1}}
\def\chibar{{\overline{\chi}}}
\def\Jbar{{\overline{J}}}
\def\qbar{{\overline{q}}}
\def\calO{{\cal O}}
\def\calE{{\cal E}}
\def\calT{{\cal T}}
\def\calM{{\cal M}}
\def\calF{{\cal F}}
\def\calY{{\cal Y}}
\def\calL{{\cal L}}
\def\rep#1{{\bf {#1}}}
\def\ie{{\it i.e.}\/}
\def\eg{{\it e.g.}\/}
\def\eleven{{(11)}}
\def\ten{{(10)}}
\def\nine{{(9)}}
\def\Ip{{\rm I'}}
\def\oneprime{{I$'$}}
\hyphenation{su-per-sym-met-ric non-su-per-sym-met-ric}
\hyphenation{space-time-super-sym-met-ric}
\hyphenation{mod-u-lar mod-u-lar--in-var-i-ant}


\def\inbar{\,\vrule height1.5ex width.4pt depth0pt}

\def\IC{\relax\hbox{$\inbar\kern-.3em{\rm C}$}}
\def\IQ{\relax\hbox{$\inbar\kern-.3em{\rm Q}$}}
\def\IR{\relax{\rm I\kern-.18em R}}
 \font\cmss=cmss10 \font\cmsss=cmss10 at 7pt
\def\IZ{\relax\ifmmode\mathchoice
 {\hbox{\cmss Z\kern-.4em Z}}{\hbox{\cmss Z\kern-.4em Z}}
 {\lower.9pt\hbox{\cmsss Z\kern-.4em Z}}
 {\lower1.2pt\hbox{\cmsss Z\kern-.4em Z}}\else{\cmss Z\kern-.4em Z}\fi}

\def\NPB#1#2#3{{\it Nucl.\ Phys.}\/ {\bf B#1} (19#2) #3}
\def\PLB#1#2#3{{\it Phys.\ Lett.}\/ {\bf B#1} (19#2) #3}
\def\PRD#1#2#3{{\it Phys.\ Rev.}\/ {\bf D#1} (19#2) #3}
\def\PRL#1#2#3{{\it Phys.\ Rev.\ Lett.}\/ {\bf #1} (19#2) #3}
\def\PRT#1#2#3{{\it Phys.\ Rep.}\/ {\bf#1} (19#2) #3}
\def\CMP#1#2#3{{\it Commun.\ Math.\ Phys.}\/ {\bf#1} (19#2) #3}
\def\MODA#1#2#3{{\it Mod.\ Phys.\ Lett.}\/ {\bf A#1} (19#2) #3}
\def\IJMP#1#2#3{{\it Int.\ J.\ Mod.\ Phys.}\/ {\bf A#1} (19#2) #3}
\def\NUVC#1#2#3{{\it Nuovo Cimento}\/ {\bf #1A} (#2) #3}
\def\etal{{\it et al.\/}}

\long\def\@caption#1[#2]#3{\par\addcontentsline{\csname
  ext@#1\endcsname}{#1}{\protect\numberline{\csname
  the#1\endcsname}{\ignorespaces #2}}\begingroup
    \small
    \@parboxrestore
    \@makecaption{\csname fnum@#1\endcsname}{\ignorespaces #3}\par
  \endgroup}
\catcode`@=12

\input epsf


\section{Introduction}
\setcounter{footnote}{0}

One of the most surprising theoretical developments of the past few years
has been the realization that the fundamental high energy scales of physics are
not immutable, and that they can be altered in the presence of 
extra spacetime dimensions.  
Specifically, it has been shown that extra spacetime dimensions have
the potential to lower the fundamental GUT scale~\cite{DDG}, the
fundamental Planck scale~\cite{ADD}, and the fundamental string 
scale~\cite{WittLykk};  indeed, each of these scales can be lowered
to potentially accessible energy scales in the multi-TeV range.
Despite their similarities, however, these scenarios have important
differences.  The GUT scale-lowering scenario of Ref.~\cite{DDG}
utilizes extra ``in-the-brane'' dimensions that are roughly of the same order   
as the (lowered) fundamental scale of the higher-dimensional theory. 
Thus, no hierarchy is introduced.
By contrast, the
Planck scale-lowering scenario of Ref.~\cite{ADD} requires extra 
``off-the-brane'' dimensions that are hierarchically larger than 
the higher-dimensional fundamental scale.
Thus, the hierarchy between the weak scale and four-dimensional
Planck scale is not explained, but rather reformulated 
in a geometric context as a new hierarchy between the weak scale 
and the scale of extra dimensions. 
Nevertheless, as discussed in Ref.~\cite{DDG},
it is possible to combine all three of these scenarios 
in a consistent way within the framework of Type~I string 
theory.  Thus, combining these scenarios, a consistent picture of 
reduced gauge and gravitational energy scales emerges.   

Recently, a new proposal~\cite{RS1} has been made for generating the
Planck-scale/weak-scale hierarchy without the use of large extra
dimensions, but rather as a result of gravity localization.  
As such, this scenario explains the apparent weakness
of gravity without recourse to large extra dimensions, and provides
an interesting alternative to the large-dimension scenario of Ref.~\cite{ADD}.
The basic idea is as follows.
As shown in Ref.~\cite{RS1}, gravity localization emerges naturally in a
D-brane context when the spacetime gravitational effects of the D-brane itself
are taken into account.  Such gravity localization arises because the
presence of the D-brane induces the spacetime metric to accrue a scale factor 
(``warp factor'') which is 
a falling exponential function of the distance along the dimension 
perpendicular to the brane.  Thus, the graviton is essentially ``bound'' 
or localized to the D-brane. 
By imagining that the Standard Model is restricted to a second D-brane whose
position is shifted relative to the first, one finds that a hierarchically small scale
factor is generated for the metric on the second brane. 
This in turn requires a rescaling of the fields 
on the Standard-Model brane, which has the net effect of generating
an exponential hierarchy between the mass scales on the Standard-Model
brane and the fundamental (higher-dimensional) mass scales.
For example, a TeV-sized electroweak scale can be generated on   
the Standard-Model brane even when this brane is shifted by only a small
amount (in Planck-scale units) from the original localization brane. 
Various generalizations and extensions of this scenario have been
considered in Refs.~\cite{Verlinde,CC,GWbulkfields,others,Kaloper};  
likewise, earlier solutions involving different ``warp factors'' can be found 
in Ref.~\cite{ovrut}.

While this scenario elegantly generates the desired hierarchy between
the Planck scale and the electroweak scale, 
certain features are left unexplained.
One important issue, for example, 
is to explain how gauge unification
might arise in such a context.  
This issue is particularly pressing for the following
reason.
In such a gravity-localized scenario, 
the presence of an exponential warp factor requires that the
Standard-Model gauge groups correspond to parallel, coincident 
D-branes.  
Indeed, because of the exponential warp factor,
even small relative displacements amongst these Standard-Model branes
could have potentially large unwanted effects.
However, for simplicity, it is natural to expect that the gauge couplings
should all take a common, unified value if their corresponding 
gauge groups arise from parallel, coincident D-branes.
Thus, in a gravity-localized compactification, one 
expects to have only a single, unified gauge coupling
for all of the Standard-Model gauge-group factors.
Moreover, this unification of gauge couplings
should {\it a priori}\/ arise directly at (or near) the electroweak 
scale, which is interpreted as the only physical scale
on the Standard-Model brane.
It then remains to explain why these gauge couplings are
experimentally measured to be different.

In this paper, we shall show that
      gauge coupling unification arises naturally in
      such scenarios as a result of the anomaly induced by
      the rescaling of the wavefunctions of the brane fields.
Specifically, we shall show that the anomaly produces a contribution
to the gauge couplings that splits them in such a way that
they appear to have emerged from a traditional high-scale logarithmic
unification, or equivalently from a low-scale power-law unification
as in Ref.~\cite{DDG}.

The success of gauge coupling unification is thus
a compelling issue in favor of such gravity-localized scenarios.
      However, we also point out a number of difficulties
      with such scenarios, including an inability to 
      accommodate the GUT scale and the electroweak scale
      simultaneously.  Moreover, we also find that brane/bulk couplings
      in this scenario are generically too small to be
      phenomenologically relevant.  
We believe that these issues are generic to the scenario of 
Ref.~\cite{RS1}, and will need to be overcome before a serious
investigation of the phenomenology of these scenarios is possible.

Motivated by our results, we then proceed to discuss a 
possible modification of this scenario which avoids some of these problems.
Our modification consists of changing the slope of the warp factor so
that the warp factor is maximized rather than minimized on the  
Standard-Model brane.   In this respect our proposal is similar
to that of Ref.~\cite{RS2}, except that we shall take the radius
of the extra dimension to be finite.
As we shall
see, gauge coupling unification is also easily accommodated in this
modified scenario, while brane/bulk couplings take more reasonable sizes and 
even neutrino masses can be accurately predicted.  Proton decay is also
sufficiently suppressed, and the scenario as a whole  
is consistent with an expanding Friedmann-like universe.
However, in such a scenario, the generation of the electroweak
scale is an important outstanding question.  We shall discuss each
of these issues in turn, and speculate on some possible resolutions
to these puzzles.

\section{The framework}
\setcounter{footnote}{0}

We begin by briefly reviewing the scenario of Ref.~\cite{RS1}.
This will also enable us to establish our physical and notational 
conventions, which differ from those of Ref.~\cite{RS1} in some
significant ways.  It will also enable us to introduce our modified
scenario with a flipped warp factor.

\subsection{General setup}

As in Ref.~\cite{RS1},
we take spacetime to be five-dimensional, with the fifth dimension
compactified on a $\IZ_2$ orbifold of radius $R$.  Thus,
the fifth dimension is essentially a line interval parametrized
by a coordinate $y$ stretching over the finite interval $0\leq y\leq \pi R$.
If required, one can then formally extend the $y$-coordinate to all real values
using the orbifold symmetry relations
\beq
           y~\approx ~y+2\pi R~,~~~~~
            y~\approx ~-y~\approx~ 2\pi R-y ~. 
\label{orbrelations}
\eeq
We also assume the presence of two D3-branes, one located at each
orbifold fixed point, as well as a bulk cosmological constant $\Lambda<0$.
In Ref.~\cite{RS1}, the D3-brane located at $y=\pi R$ is presumed
to be the one containing the Standard Model.
Thus, just as in Ref.~\cite{RS1}, the classical action 
describing this situation is given by
\beqn
      S &=& \int d^4 x \,\biggl\lbrace
         \sqrt{-g^{(0)}} \,(\calL^{(0)} - V^{(0)})
         ~+~ \sqrt{-g^{(\pi R)}} \,(\calL^{(\pi R)} - V^{(\pi R)}) \nonumber\\
            && ~~~~~~~~~~~~+~ \int_{-\pi R}^{\pi R} dy \, \sqrt{-G} 
             (-\Lambda + 2 M^3 R^{(5)}) \biggr\rbrace
\label{action}
\eeqn
where $G$ is the bulk five-dimensional metric (with negative determinant); 
$R^{(5)}$ is the five-dimensional curvature derived from $G$;
$g^{(0)}$ and $g^{(\pi R)}$ are
the induced four-dimensional metrics on the D3-branes; 
and $M$ is the fundamental
five-dimensional mass scale in the bulk.
The goal is then to solve the corresponding Einstein field equations
for the five-dimensional bulk metric $G$.

If, as in Ref.~\cite{RS1}, we take a trial solution of the form
\beq
      ds^2 ~=~ e^{-2\sigma(y)} \eta_{\mu\nu} dx^\mu dx^\nu ~+~ dy^2~,
\label{trial}
\eeq
then it is straightforward to derive two differential equations
for the unknown function $\sigma(y)$.  
It turns out~\cite{RS1} that these differential equations do not have solution unless
$V^{(0)}$, $V^{(\pi R)}$, and $\Lambda$ are related according to
\beq
    V^{(0)} = -V^{(\pi R)} = 24 M^3 k~,~~~~~~~
         \Lambda = - 24 M^3 k^2~
\label{relations}
\eeq
where $k$ is an arbitrary mass scale.
In terms of $k$, the two differential equations then take the form~\cite{RS1}
\beq
     \left(d\sigma\over dy\right)^2 = k^2~,~~~~~ {d^2 \sigma\over dy^2} ~=~
          - {2k^2\over \Lambda} \left\lbrack V^{(0)} \delta(y) +
                V^{(\pi R)} \delta(y-\pi R)\right\rbrack~.
\label{diffeqs}
\eeq

The next step is to solve these differential equations for $\sigma(y)$.
In Ref.~\cite{RS1}, the solution is taken to be 
\beq
       \sigma(y) ~=~ \cases{
              ky & $0\leq y\leq \pi R$ \cr
              k(2\pi R-y) & $\pi R \leq y \leq 2\pi R$~. \cr}
\label{RSsoln}
\eeq
Given this, one can then calculate the effective four-dimensional Planck 
mass $M_{\rm Planck}$ directly from the five-dimensional curvature 
term in (\ref{action}) 
by inserting (\ref{RSsoln}) into the metric (\ref{trial})
and integrating over the fifth dimension.
This yields the result~\cite{RS1}        
\beq
      M_{\rm Planck}^2 ~=~ 2 M^3 \int_0^{\pi R} dy \,e^{-2 ky} ~=~
         {M^3\over k} \left( 1- e^{-2\pi k R}\right)~,
\label{RSplanckmass}
\eeq
on the basis of which the authors of Ref.~\cite{RS1} are compelled
to choose $M$ and $k$ to be of roughly the same order 
of magnitude as $M_{\rm Planck}$.

\subsection{Orbifold symmetries}

In this paper, we wish to consider two distinct scenarios.
The first will be the above scenario, with $k$ taken to be a positive
quantity, while the second will be the scenario in which $k$ is taken
to be negative.  
However, 
taking $k$ negative in the above solution
leads to certain subtleties which
are ultimately spurious, and which disguise the physics
we hope to discuss. 
Therefore, before proceeding further,  we shall
find it useful to make one additional change in the solution
(\ref{RSsoln}).

Of course, (\ref{RSsoln}) is
a perfectly valid solution to the differential equations (\ref{diffeqs}).
Moreover, as noted in Ref.~\cite{RS1}, one is always free to add an arbitrary
$y$-independent constant $\sigma_0$ 
to the solution in (\ref{RSsoln}),
for this simply amounts to an overall constant rescaling of the four-dimensional
metric which in turn amounts to a rescaling of the four-dimensional
spacetime coordinates.  Such a rescaling therefore has no net
physical effect,  
and in particular does not change the values of physical
quantities such as $M_{\rm Planck}$.
Indeed, the choice of $\sigma_0$ is analogous to a gauge choice,
since the physics is ultimately invariant under changes in $\sigma_0$, and all 
$\sigma_0$-dependence ultimately cancels in calculations of physical quantites.
However, just as with gauge theories,
certain choices of $\sigma_0$ can make certain features of the calculation
more transparent than others.

Although $\sigma_0=0$ is the choice taken in Ref.~\cite{RS1}, 
in this paper we shall use a different choice for $\sigma_0$.
Our choice is motivated by the fact that (\ref{RSsoln}) by itself does not
make manifest the full orbifold
symmetries of the theory.
In order to see this, let us consider the following simultaneous
transformations
\beq
           \cases{ y \to y + \pi R \cr
                   k \to -k ~.\cr}
\label{perspective}
\eeq
The first transformation simply amounts to a translation of
the $y$-coordinate
by half of the full length of the orbifold.
Using the orbifold relations (\ref{orbrelations}), we see
that $y=0$ is mapped to $y=\pi R$, while
$y=\pi R$ is mapped to $y=2\pi R$, which is equivalent
to $y=0$.
Thus, the mapping $y\to y+\pi R$ simply amounts to 
exchanging the positions of the two D3-branes. 
Likewise, from (\ref{relations}), it is clear that changing $k\to -k$
also amounts to exchanging the role of the two branes.
Thus, we see that the simultaneous transformations given in (\ref{perspective})
should have no net effect, and should therefore be a symmetry of the theory.
Unfortunately, the solution given in (\ref{RSsoln}) does not manifest
this symmetry.
Indeed, rather than remaining invariant under (\ref{perspective}),
we see that
\beq
     \sigma(y + \pi R, -k ) ~=~  \sigma(y,k) - \pi k R~. 
\eeq
Thus, the solution for $\sigma(y)$ given in 
(\ref{RSsoln}) forces us to perform an overall shift in the
absolute value of $\sigma$
each time we perform the symmetry operations (\ref{perspective}).

Even though such an overall shift is unphysical,
it is awkward to deal with a solution that 
requires compensating rescalings under orbifold symmetry shifts.
This will be particularly troublesome when we choose to consider
$k$ to be negative rather than positive.
In order to remedy this situation, let us therefore consider a slightly more 
general solution of the form
\beq
        \sigma(y) ~=~ k y + k \sigma_0 ~,~~~~~~ 0\leq y \leq \pi R~
\label{presoln}
\eeq
where $\sigma_0$ is an arbitrary constant to be determined.
By the orbifold relations (\ref{orbrelations}),
this implies 
        $\sigma(y) = k (2\pi R-y) + k \sigma_0$
in the range $\pi R \leq y \leq 2\pi R$.
Indeed, since $\sigma_0$ is presumed independent of $y$, this new solution
will also satisfy the same differential equations 
(\ref{diffeqs}).
Demanding invariance under (\ref{perspective}), we can then solve
for $\sigma_0$, yielding the unique result $\sigma_0 = - \half \pi R$.
We therefore conclude that the modified solution
\beq
           \sigma(y) ~=~ \cases{
                 k y  - \half \pi k R & $0\leq y \leq \pi R$\cr
                 k (2\pi R- y) - \half \pi k R & $\pi R\leq y \leq 2\pi R$\cr}
\label{DDGsoln}
\eeq
not only satisfies the differential equations (\ref{diffeqs}), but also
exhibits the required invariance under the full orbifold 
symmetry relations (\ref{perspective}).

Even though
the solution (\ref{DDGsoln})  
is not physically different from that in (\ref{RSsoln}), 
the passage from (\ref{RSsoln}) to (\ref{DDGsoln}) does induce a 
change in perspective which can often prove useful.  
As an example of this,
let us repeat
the calculation of the effective four-dimensional
reduced Planck mass $M_{\rm Planck}$ using the shifted solution
(\ref{DDGsoln}) for the five-dimensional metric.
Following the same procedure as in (\ref{RSplanckmass}), we now obtain
the result
\beq
      M_{\rm Planck}^2 ~=~ 2 M^3 \int_0^{\pi R} dy 
             \,e^{-2 ky  + \pi k R} ~=~  {2 M^3\over k}\, \sinh \left( \pi kR\right)~.  
\label{DDGplanckmass}
\eeq
Note that, unlike (\ref{RSplanckmass}), this result 
for $M_{\rm Planck}$ is now invariant under $k\to -k$. 
It is desirable to have this invariance under $k\to -k$
because we have integrated over the 
full fifth dimension when calculating $M_{\rm Planck}$, 
and therefore we are not distinguishing which D3-brane is located in
which position.  Thus, having $M_{\rm Planck}$ be invariant
under $k\to -k$ more closely reflects the inherent symmetries
of the system.

Although (\ref{RSplanckmass}) and (\ref{DDGplanckmass}) differ
by only an overall scale factor, their physical interpretations
are different.  Whereas previously the authors of Ref.~\cite{RS1}
were forced to take $M$ and $k$ to be near $M_{\rm Planck}$, we now 
see from (\ref{DDGplanckmass}) that $M_{\rm Planck}$ can take its 
correct apparent four-dimensional value even when
$M$ and $k$ are taken substantially smaller.  
This arises because of our choice of taking $\sigma_0\not=0$, since
$M$ and $k$ are ultimately $\sigma_0$-dependent quantities.
For example, taking $M=k=10$ TeV as a typical small reference fundamental
mass scale, we find that the apparent value of $M_{\rm Planck}$ 
can be accommodated simply by taking $k R \approx 21$.
Alternatively, if we choose $M\approx 10^{10}$ GeV
(as might be preferred on the basis of solving the gauge hierarchy
problem as in Ref.~\cite{RS1}), we would require $kR\approx 12$. 
However, in order for our classical gravity treatment of the
brane system to be consistent, we actually must require that
$k/M\ll 1$;  otherwise the
curvature terms in the effective
Lagrangian cannot be neglected.
Depending on the values of the ratio $k/M$,
the above estimates for $kR$ will change as well.
We shall discuss this point in subsequent sections.

Thus, we see that we can generate a sufficiently high Planck scale,
even with fundamental bulk mass scales $M$ and $k$ near the
electroweak range,
simply by taking the fifth dimension to have a radius that
is also near the electroweak range.
It is easy to interpet this result physically.
The apparent four-dimensional Planck mass can be large
(even when the fundamental physical scales $M$ and $k$ are small) because
the effective {\it volume}\/ $L_{\rm eff}$ of 
the fifth dimension
can be large
(even though the compactification radius $R$ is relatively small).
Specifically, we find
\beq
        M^2_{\rm Planck} ~=~ M^3 \,L_{\rm eff}~,
\eeq
where
\beq
    L_{\rm eff} ~=~ 2 \int_0^{\pi R} dy \,e^{-2 ky  + \pi k R} ~=~ 
             2\, k^{-1} \, \sinh(\pi k R)~ \gg~ R~.
\label{Leff}
\eeq
Thus, the ``warp factor'' of the bulk metric 
has enhanced the small radius into a large volume, thereby enabling
the effective four-dimensional Planck mass $M_{\rm Planck}$ to 
be hierarchically larger than the five-dimensional fundamental scale $M$.

Of course, we stress again that there is ultimately no physical distinction
between the results in (\ref{RSplanckmass}) and (\ref{DDGplanckmass}).
Rather, the change in metric from (\ref{RSsoln}) to (\ref{DDGsoln}) has
merely red-shifted our definitions for the mass scales $(M,k)$ in the bulk
from the four-dimensional Planck scale (as in Ref.~\cite{RS1}) to 
to scales that are much lower.
Indeed, $M$ and $k$ are ultimately unphysical quantities, depending on
our particular choice for the overall absolute scale factor of the five-dimensional
metric.  Only $M_{\rm Planck}$, as well as other mass scales on the Standard-Model
brane, are truly physical.
Thus, despite recent erroneous claims in the literature~\cite{Li}, there
is no physical distinction between (\ref{RSplanckmass}) and (\ref{DDGplanckmass}) 
or any other version of this result which involves
an overall constant rescaling of the five-dimensional metric. 
However, as we have seen,
the conventions we have established here more naturally reflect
the symmetries in the system.

\subsection{Two scenarios:  $k>0$ and $k<0$}

We shall be considering two
distinct scenarios in this paper.
The first, as described above, corresponds to taking $k$ positive.
The five-dimensional metric is given in (\ref{DDGsoln}), and the
Planck mass is given in (\ref{DDGplanckmass}).  We shall discuss the
appropriate values of $M$, $k$, and $kR$ in subsequent sections, but
it is natural to think of $M$ and $k$ as being substantially below the
usual Planck scale (perhaps even as low as the TeV-range), and 
$kR\approx {\cal O}(10-20)$.  We stress, however, that it is generally
necessary to have a small hierarchy $k/M\ll 1$ 
in order to justify our classical treatment 
wherein we neglected the
curvature terms in the effective Lagrangian.
This can change the appropriate values of $M$ and $kR$.

By contrast, the second scenario that we shall discuss corresponds
to taking $k$ negative.  Note that it is immediately apparent from 
the above discussion that there is nothing that compels
us to consider $k$ to be a positive quantity.
Indeed, the above solutions remain completely valid
if we consider $k$ to be negative rather than positive.
Moreover, 
as we have seen, the transformation $k\to -k$ is {\it not}\/ a
symmetry of the theory unless simultaneously accompanied by the
shift $y\to y+\pi R$
(which we will {\it not}\/ do).  Thus, the solution with $k<0$
is physically distinct from the solution with $k>0$.
As evident from (\ref{relations}), the replacement $k\to -k$
exchanges the signs of the brane potentials, so that the brane at $y=\pi R$
(which we will continue to identify as the brane containing the
Standard Model) becomes the one with {\it positive}\/ 
potential, \ie, $V^{(\pi R)} > 0$.
In other words, in this scenario
the ``warp factor'' $e^{-2\sigma(y)}$ is {\it maximized}\/ rather than
minimized on the Standard-Model brane  at $y=\pi R$.
In this respect, the negative-$k$ scenario 
resembles the toy model of Ref.~\cite{RS2}, whose purpose was
to illustrate the decoupling effects of an infinitely
large extra dimension.  Unlike Ref.~\cite{RS2}, however,
in this paper we shall treat the length of the extra dimension as finite,
and study the phenomenological properties of the resulting
brane configuration.
In a cosmological context, we also remark that the negative-$k$ solution,
with its positive-energy Standard-Model D3-brane, would also be consistent
with a Friedmann-like expanding universe~\cite{CC}.

Note that this change $k\to -k$
merely changes the sign of the brane potentials, but cannot
alter the value of quantities such as the fundamental Planck mass.
It is for this reason that we have taken the care to establish our
conventions such that $M_{\rm Planck}$ is manifestly invariant
under $k\to -k$.  If we had remained with the original conventions
in (\ref{RSsoln}), the change $k\to -k$ would have
intrinsically involved a spurious overall blue-shifting
that we would have had to subsequently disentangle from the
effects of having changed the signs of the brane potentials.
Indeed, this type of spurious blue-shift has led 
previous authors~\cite{Li} to erroneous conclusions,
so we cannot emphasize this point strongly enough. 

Thus, in the negative-$k$ solution, we shall work with a bulk
metric of the form
\beq
       ds^2 ~=~ e^{2 \hat k y - \pi \hat k R} \, \eta_{\mu\nu} dx^\mu dx^\nu
             ~+~ dy^2 ~,~~~~~ 0\leq y\leq \pi R~,
\label{DDGmetricnegative}
\eeq
where we have defined $\hat k\equiv -k>0$.
The corresponding effective four-dimensional Planck mass is given by
\beq
       M^2_{\rm Planck} ~=~  {2 M^3 \over \hat k} \,\sinh  \left(\pi \hat k R
                          \right)~,
\label{MPlanckDDG}
\eeq
and we shall discuss the particular values of $M$, $\hat k$, and $\hat kR$
as we proceed.
For example, if we take $M=10$ TeV and
demand the existence of a small 
hierarchy $\hat k/M \approx 10^{-4}$,
we then find that the apparent value of $M_{\rm Planck}$
can be accommodated with
\beq
               \hat kR ~\approx ~ 18~.
\label{radius}
\eeq
Of course, 
depending on the chosen values of $M$ and $\hat k$,
other values for $\hat k R$ remain possible. 
In all cases we shall continue to identify the D3-brane
at $y=\pi R$ as the brane containing the Standard Model.

\section{Gauge coupling unification} 
\setcounter{footnote}{0}

We begin by discussing the issue of gauge coupling unification on the 
Standard-Model brane, in both the positive- and negative-$k$ scenarios. 
As discussed in the Introduction,
the very nature of the gravity-localized framework with
its rapidly changing warp factor  
requires that the D-branes that comprise the Standard-Model
gauge-group factors be parallel and essentially coincident
at $y=\pi R$. 
This implies, in the most straightforward string embeddings,
that the gauge couplings corresponding to the different 
gauge-group factors be essentially equal 
to each other at the fundamental energy scale $M$ (which we
are imagining to be near the electroweak scale).
How then can we explain the different observed values of
the gauge couplings?  

One idea, of course, is to make use of the mechanism advanced
in Ref.~\cite{DDG} --- namely, to introduce additional ``in-the-brane''
spacetime dimensions and invoke power-law running for the gauge couplings
as a mechanism producing an accelerated unification.  
This would then require the introduction of another free parameter, 
namely the radius of the extra ``in-the-brane'' dimension, and
would explain the observed difference in low-energy gauge couplings
as a one-loop effect.
While in principle this mechanism works even in the gravity-localized
context, in this paper we shall consider something different.
Essentially, we shall utilize the extra ``{\it off}\/-the-brane'' 
dimension involved in gravity localization 
in order to achieve gauge coupling unification 
directly, 
  {\it without}\/ the need for renormalization-group running.
This would then be an economical explanation of the observed
gauge couplings within the gravity-localized framework. 

Ordinarily, it might not seem
possible for
extra dimensions 
perpendicular to the Standard-Model
brane 
to affect the gauge couplings on the brane in a group-dependent manner.   
However, in theories involving a warp factor, this is 
precisely what occurs.
Because of
the non-trivial warp factor, it 
is necessary to rescale the fields on the Standard-Model
brane in order to give them a canonical normalization.
However, this rescaling is generally anomalous, and 
induces a shift in the gauge couplings on the Standard-Model brane.
Remarkably, we shall find that this shift in the gauge couplings
can precisely account for the observed difference in the 
gauge couplings at the electroweak scale, even if we assume
that their ``bare'' values are universally coupled to the 
dilaton and hence unified.
Thus, in this framework, we {\it automatically}\/ have
a single, unified tree-level gauge coupling on the Standard-Model
brane,
and it is only an additional anomaly contribution, induced by
the warp-factor rescaling, that
gives these gauge couplings the different apparent values they
are measured to have.
We shall therefore refer to this phenomenon as ``anomaly-induced'' 
gauge coupling unification.

\subsection{The $k<0$ scenario}

Although our ``anomaly-induced'' mechanism for gauge coupling unification 
is general and applies to both the positive-$k$ and negative-$k$ solutions, 
it will prove simpler to first consider the negative-$k$ solution. 
Accordingly, let us begin with the five-dimensional bulk metric
given in (\ref{DDGmetricnegative}), where we assume that the Standard-Model
brane is located at $y=\pi R$.
This then induces a four-dimensional 
metric on the Standard-Model D3-brane
given by
\beq
            ds^2 ~=~ e^{\pi \hat k R} \, \eta_{\mu\nu} dx^\mu dx^\nu~, 
\label{SMbranemetric}
\eeq            
whereupon the kinetic-energy terms of the corresponding 
D3-brane Lagrangian will take the form
\beq
   {\cal L} ~=~  \int d^4 x \, \left( e^{ \pi \hat k R}\,|D_\mu \Phi |^2 ~+~ 
       e^{3 \pi \hat k R/2} \, \bar{\Psi} i \gamma^\mu D_\mu \Psi ~-~ 
       {1\over 4 g_i^2} {\rm Tr}\, F_{\mu\nu, i}^2   \right)~.
\label{lag}
\eeq
Here $\Phi$, $\Psi$, and $A_\mu$ respectively represent 
a complex scalar, Dirac fermion, and gauge field. 
In order to canonically normalize the kinetic-energy terms in (\ref{lag}), 
each of these fields must be Weyl-rescaled by an amount
\beq
   \Phi \rightarrow e^\Lambda \,\Phi       ~,~~~~~~
   \Psi \rightarrow e^{3\Lambda/2} \, \Psi ~,~~~~~~
     A_\mu\rightarrow A_\mu
\label{Weylrescaling}
\eeq
where $\Lambda= -\pi \hat k R/2$.
However, while the Lagrangian is classically invariant
under such a Weyl-rescaling, it is well-known
that this symmetry is anomalous at the quantum level. 
In other words, the quantum functional-integral measure does not respect
this rescaling symmetry, 
and the resulting Jacobian determinant leads to an extra term in the
Lagrangian of the form\footnote{
          In order to see why the Weyl anomaly is proportional to the
          beta-functions (which are usually associated with the breaking
          of invariance under {\it scale}\/ transformations),
         let us assume that we start with an action in curved space 
         which is invariant under the Einstein and Weyl transformations
         \beqn
              g'_{\mu \nu} = e^{-2 \Lambda} g_{\mu \nu} ~, ~~~
             \Phi' =e^{\Lambda} \Phi ~,~~~ 
             \Psi' =e^{3 \Lambda / 2} \Psi ~,~~~   
              A_{\mu}' = A_{\mu} ~.
         \label{4.1}
         \eeqn
         When restricted to flat space ($g_{\mu \nu}=\eta_{\mu \nu}$), 
         such an action is invariant under the fifteen-parameter 
          four-dimensional conformal group, which  contains, 
         in particular, the scale transformations.
         A simple way to see that Weyl transformations imply scale
         transformations is to start with (\ref{4.1}) and find an equivalent
         transformation in flat space such that $d^4 x \sqrt{\det g'} = d^4 x'$. 
         This yields $x'=e^{-\Lambda}x$, and when substituted into the Lagrangian
         this implies the transformations
         \beqn
             x' = e^{- \Lambda} x ~,~~~ \Phi'(x') =e^{\Lambda} \Phi (x) ~,~~~ 
            \Psi' (x') =e^{3 \Lambda / 2} \Psi (x) ~,~~~ 
              A_{\mu}' (x') =e^{\Lambda} A_{\mu} (x) ~.
         \label{4.2}
         \eeqn
         These are precisely the scale transformations.  In particular,
         under $y\rightarrow y+\lambda$, we find using (\ref{DDGsoln}) that 
         $x\rightarrow e^{{\hat k}\lambda}x$. }
\beq
    \delta {\cal L}_{\rm anomaly} ~=~  
       \Lambda \, \sum_i  {\beta(g_i)\over 2 g_i^3}
        {\rm Tr}\, F_{\mu\nu,i}^2~ ~+~ ...
\label{weylformula}
\eeq
where the beta-functions $\beta(g_i)$ are defined in terms of the
one-loop beta-function coefficients $b_i$ via
\beq
        \beta(g_i) ~=~ {b_i\over 16\pi^2} \, g_i^3~.
\label{betafunction}
\eeq
Note that in writing (\ref{weylformula}) and (\ref{betafunction}), we have
chosen a sign convention such that an asymptotically free theory
has $b_i <0$.  We shall also assume that the theory on the brane 
is the Minimal Supersymmetric Standard Model (MSSM) 
or some other theory that by itself would be consistent with
a conventional high-scale logarithmic unification.
Thus, performing the Weyl rescaling (\ref{Weylrescaling}), we find that
the kinetic-energy terms on the Standard-Model brane 
will all have the proper canonical normalizations,
but the kinetic-energy terms for the gauge fields 
now take the form
\beq
    {\cal L}+\delta {\cal L}_{\rm anomaly}
           ~=~  - {\textstyle {1\over 4}} \sum_i 
                 \int d^4 x   
     \left( {1\over g_i^2} + {b_i \over 16 \pi} \, \hat k R\right) 
         \,{\rm Tr}\, F_{\mu\nu,i}^2 ~ +  ...~.
\eeq
Assuming that the ``bare'' couplings $g_i$ are all unified at a
common value $g_U$ (as might be determined in a string framework
through the vacuum expectation value of the dilaton), 
this enables us to identify the physical couplings as
\beq
     {1\over g_i^2}\Biggl|_{\rm phys}~\equiv~ 
         {1\over g_U^2} + {b_i \over 16 \pi} \,\hat k R~.
\label{powerlaw}
\eeq

At first glance, this result might not seem to be satisfactory,
for we see that the scale anomaly has failed to generate the expected
logarithmic term that would be required for a traditional
unification of the gauge couplings.  Nevertheless, it turns out
that this result still leads to a consistent unification.
At a mathematical level, 
the reason for this coincidence is that
the result (\ref{powerlaw}) is 
formally identical to the power-law ``accelerated'' unification that 
was already previously discussed in Ref.~\cite{DDG}.  
Indeed, in the language of Ref.~\cite{DDG},
we see that the anomaly has precisely reproduced 
the case with $\delta=1$ and $\tilde b_i = b_i$. 
The fact that $\tilde b_i =b_i$ then guarantees that the
unification of gauge couplings in this scenario is exactly
as precise as it would have been in the MSSM.   
Indeed, taking $g_U \approx 0.7$ (which is the usual 
unified coupling in the MSSM) and $kR\approx 18$, we find 
that we obtain exactly the same couplings $g_i$ at the fundamental 
scale $M=10$  TeV as we would have obtained in the MSSM!
Thus, we see that the scale anomaly yields precisely the correct
``threshold'' corrections that split the observed gauge couplings
by an amount that simulates the effects of a conventional logarithmic
running over fourteen orders of magnitude.
In this respect, 
our anomaly-induced gauge coupling unification is similar in
spirit to the ``mirage unification'' proposals of Ref.~\cite{Ibanez}.
 
Although the result (\ref{powerlaw}) is formally identical
to the power-law unification discussed in Ref.~\cite{DDG},
we stress that (\ref{powerlaw}) is {\it not}\/ to be interpreted
as resulting from a higher-dimensional running of gauge couplings.
Rather, the second term in (\ref{powerlaw}) arises from a quantum
anomaly, and does not involve the contributions of any Standard-Model 
Kaluza-Klein states.
Nevertheless, it is remarkable that the anomaly has generated
precisely the term which can unify the gauge couplings,
even without Kaluza-Klein excitations for the Standard-Model gauge
or matter fields.
This ultimately arises because the anomaly term gives contributions
that are proportional to the one-loop beta-function coefficients
of the theory living on the Standard-Model D3-brane.  

It is also possible to understand this result as a blue-shifting 
effect on the Standard-Model D3-brane.
To see this, let us imagine running the gauge couplings
from the fundamental scale $M$ down to their observed
values at the $Z$-scale $M_Z$.
Adding this one-loop running contribution to our result (\ref{powerlaw})
then yields
\beq
     {1\over g_i^2(M_Z)}~=~ 
         {1\over g_U^2}  ~-~ {b_i\over 8\pi^2 } \ln\left( {M_Z\over
         M} \right) 
           ~+~ {b_i \over 16 \pi} \, \hat k R~,
\eeq
which can be rewritten in the form
\beq
     {1\over g_i^2(M_Z)}  ~=~
    {1\over g_U^2}  ~-~ {b_i\over 8\pi^2 } \ln\left( 
                   {M_Z\over M e^{\pi \hat k R/2}}\right)~.
\label{MGUTprime}
\eeq
Consequently, identifying $M_{\rm GUT} \equiv  M e^{\pi \hat k R/2}$,
we see that the usual GUT scale $M_{\rm GUT} \approx 10^{16}$ GeV 
can be obtained from the fundamental scale $M \approx 10$ TeV via the
 {\it blue-shifting}\/ induced by the warp factor in the metric.
Indeed, as stated above, all that is required is a value 
${\hat k}R \approx 18$,
which is the same value as required in order to generate the correct
Planck mass in (\ref{MPlanckDDG}).
Thus, through this blue-shifting effect, 
we see that a unified gauge coupling at the low scale $M=10$ TeV
is consistent with the observed values of the low-energy gauge 
couplings.
Of course, in the above equations it is not really necessary to identify
the ultraviolet cutoff with the fundamental physical scale $M$ in the bulk.
However, without knowing the full underlying theory, 
it is natural to identify these two scales for simplicity, 
and we shall continue to do so in what follows.

One interesting consequence of (\ref{powerlaw}) is that the radius
has a critical value $R^\ast$ above which the asymptotically free
gauge couplings on the Standard-Model brane become large
and ultimately diverge.  In the case of the $SU(3)$ coupling, this 
occurs at the critical value
\beq  
    {\hat k} R^\ast ~=~ {16 \pi \over |b_3| \,g_3^2 (M)} ~ . 
\label{4.3}   
\eeq
Assuming the MSSM matter content and the unified gauge coupling
$g_3 (M) = g_U\approx 0.7$, we find the critical radius
${\hat k} R^\ast \approx 30$.
Note that similar phenomena also appear in Type~I strings~\cite{AS} 
or in M-theory~\cite{Witten}.
Of course, it is natural to expect the appearance of a maximum
critical radius in any case, since an increase in the radius implies
an increase in the corresponding warp factors, which ultimately
throws part of the brane/bulk system into a non-perturbative regime.  

Given that we have chosen to identify the scale $M$ in (\ref{MGUTprime})
with the fundamental bulk mass scale, we now see that there are
two simultaneous equations that fix
the Planck and GUT scales in the negative-$k$ scenario:
\beq
          \cases{ 
     M_{\rm Planck}^2 \,\approx\, (M^3 /\hat k) \,e^{\pi \hat k R} & \cr
     M_{\rm GUT} \,=\, M \,e^{\pi \hat k R/2}~. & \cr}
\eeq
Eliminating $R$, we obtain
\beq
            M_{\rm Planck} ~=~ 
          \sqrt{M\over \hat k} \, M_{\rm GUT}~.
\label{GUTPlanck}
\eeq
This equation directly relates the fixed physical scales 
$M_{\rm GUT}$ and $M_{\rm Planck}$ to the ratio $M/\hat k$, and holds 
generically in any such gravity-localized scenario regardless of the overall
scale factor for the five-dimensional bulk metric. 
Thus,
in order to successfully reproduce\footnote{
  In passing, we remark that this also provides another solution
  to a long-standing problem~\cite{review} in the phenomenology 
  of string theory, namely to find a way of reconciling the 
  difference between $M_{\rm GUT}$ and $M_{\rm Planck}$.}
the values of $M_{\rm Planck}$ as well as
$M_{\rm GUT}$ in the negative-$k$ scenario, 
we see that we must introduce a small 
hierarchy $M/\hat k\approx 10^4$.  This justifies our earlier choice in
deriving (\ref{radius}).  
Fortunately, this small hierarchy 
is also compatible with the restriction $\hat k \ll M$ that
permitted us to trust the classical
gravity approximation in Sect.~2 
wherein we neglected the 
curvature $R^2$ terms in the effective five-dimensional Lagrangian.
 
Note that the above results can be easily generalized 
to include the running of  other parameters in the Lagrangian
of the Standard-Model D3-brane. 
In order to be specific, let us consider a theory containing 
gauge fields, fermions $\Psi$, and complex scalars $\Phi$, with a Lagrangian
\beq
  {\cal L} ~=~ \int d^4 x\, \left\lbrack
     -{1 \over 4g^2} {\rm Tr} F_{\mu \nu}^2 + {\bar \Psi}
  (i\gamma^{\mu} D_{\mu} -m_\psi) \Psi + |D_{\mu} \Phi|^2 - 
   m_\phi^2 \Phi^\dagger \Phi - {\lambda \over 4}
  (\Phi^\dagger \Phi)^2 + ...\right\rbrack
\label{g1}
\eeq
where the ellipses denote other terms irrelevant for the present discussion.
Under the Weyl rescaling (\ref{Weylrescaling}), the mass terms explicitly
break the classical scale invariance and the path-integral measure
is anomalous. This leads to an additional anomaly-generated contribution to 
the Lagrangian given by
\beqn
    \delta {\cal L}_{\rm anomaly} &=& \Lambda \left\lbrack
     {\beta(g) \over 2 g^3} \,{\rm Tr}\, F_{\mu\nu}^2 + 
       m_\psi (1 + \gamma_{m_\psi}) {\bar \Psi} \Psi + 
       m_\phi^2 (2+\gamma_{m_\phi^2}) \Phi^\dagger \Phi - 
      {1 \over 4} \beta (\lambda) (\Phi^\dagger \Phi)^2  
       \right\rbrack \nonumber\\
    &=& \Lambda \mu {\partial \over \partial \mu} {\cal L}  ~,
\label{g2}
\eeqn
where $\gamma_{m_\psi}$ (respectively $\gamma_{m_\phi}$) 
is the anomalous dimension of $\Psi$ (respectively $\Phi$).
Note that in the last line, the derivative acts only on the parameters
$g(\mu)$, $m(\mu)$, and $\lambda(\mu)$, which depend on the
renormalization scale $\mu$ according to
\beqn
 && \mu {\partial g \over \partial \mu} = \beta (g) ~, ~~~~~ 
 \mu {\partial \lambda \over \partial \mu} = \beta (\lambda) ~,~~~~~ \nonumber\\
 && \mu {\partial m_{\psi} \over \partial \mu} = -m_{\psi} 
 (1 + \gamma_{m_{\psi}}) ~,~~~~~ 
 \mu {\partial m_{\phi}^2 \over \partial \mu} = -m_{\phi}^2 
  (2 + \gamma_{m_{\phi^2}}) ~ .  
\label{4.4}
\eeqn
Thus, for every renormalized parameter $X \equiv (g,m,\lambda,...)$  in
the Standard-Model Lagrangian, we can write the equation
\beq
 \mu {\partial X \over \partial \mu} ~=~ {\partial X \over \partial \Lambda} ~.
\label{g3}
\eeq
This yields the solution  $X = X (\mu e^{\Lambda})$,
where the functional dependence of $X$ is fixed by the usual 
renormalization-group equations.   
Since $\mu$ always appears in the combination $\mu / M$, 
this explains why the 
apparent ultraviolet cutoff for 
the gauge coupling running is red-shifted from $M_{\rm GUT}$ to 
$M_{\rm GUT} e^{-\Lambda}$.
This also demonstrates that the same phenomenon actually occurs for {\it all}\/ 
running quantities in the Standard-Model Lagrangian. 

This last result is highly non-trivial.  Of course, it is clear
that the rescaling induced by the metric produces
corresponding rescaling of the bare (classical) mass parameters in our
D3-brane Standard-Model Lagrangian.
However, the unification scale $M_{\rm GUT}$ is
not a classical mass scale, but rather the result of a one-loop
quantum running.  Nevertheless, 
we have shown that 
even this {\it quantum}\/ mass scale is rescaled, thanks
to the rescaling {\it anomaly}\/ that automatically accompanies
the classical coordinate- (or field-) rescaling.
Thus, at both the classical {\it and}\/ quantum levels,
the rescalings induced by the metric have the net effect
of successfully rescaling {\it all}\/ of the mass scales on the brane.

As an example of this, let us consider the running of the top Yukawa
coupling $y_t$ in the MSSM.
For simplicity, we shall keep only the strong coupling $g_3$ 
in the renormalization group
equations and take $g_3$ to be fixed.
By integrating the renormalization-group equations, we find the solution
\beq
  {1 \over y_t^2 (\mu)} ~=~ {9 \over 8 g_3^2} \left\lbrace 
  1 - \exp\left(-{g_3^2 {\hat k} R \over 3 \pi} \right) \left[ 
  1-{8g_3^2 \over 9 y_{t_0}^2} \right]
  \left({\mu \over M}\right)^{2g_3^2 /( 3 \pi^2)} \right\rbrace ~,
\label{4.5}
\eeq
where $y_{t_0}$ is the initial value that the Yukawa coupling would 
have at the scale $M e^{\pi {\hat k} R/2}$.
Note that the new exponential factor  $\exp(-g_3^2 {\hat k} R / (3 \pi))$
is small for radii below the critical value (\ref{4.3}).
Therefore, no further constraint on the critical radius arises from 
considerations of the Yukawa couplings.

We conclude, then, that the gravity-localized scenario
with $k<0$  
automatically leads to gauge unification as a result of
the anomalous rescaling induced by the gravity-localizing 
warp factor in the spacetime metric.
This unification occurs directly at tree-level (as a result of
a ``threshold'' effect induced by the one-loop anomaly term), 
and utilizes the same relatively small extra dimensions that already 
generate the gauge hierarchy.
Despite the superficial similarity to the proposal in Ref.~\cite{DDG},
we stress that this anomaly-induced unification
scenario does not rely on a quantum-mechanical
power-law ``running'' of any sort, and in particular does not
involve Kaluza-Klein excitations for any of the Standard-Model
gauge or matter fields.  Indeed, the Standard Model continues
to be restricted to a single set of D-branes (which may therefore 
be taken to be three-branes).  Thus, in this scenario, we see
that we can simultaneously explain the weakness of gravity 
  {\it as well as}\/ the unification
of the gauge forces, all as the result of a single extra spacetime
dimension whose size is relatively close to the fundamental
physical scales in the theory.
Moreover, this occurs without large mass scales on the brane
or in the bulk.
On the other hand, as we shall see, this scenario has 
difficulty explaining the origin of the electroweak symmetry-breaking 
scale.  We shall
discuss this issue in Sect.~4.

\subsection{The $k>0$ scenario}

The above mechanism for anomaly-induced gauge coupling
unification also applies to the positive-$k$
scenario.  We simply algebraically replace
$\hat k$ in the above expressions with $-k$, and then
consider $k$ to be a positive quantity.
We thus find the relation
\beq
              M_{\rm GUT} ~=~ M e^{- \pi k R/2}
\label{gutresult}
\eeq
which must be satisfied in conjunction with the Planck mass
relation given in (\ref{DDGplanckmass}).
Eliminating $R$, we find that these equations together imply 
the relation
\beq
         M_{\rm Planck}~=~ M_{\rm GUT} \,\sqrt{M\over k} \, 
                       e^{\pi k R}~.
\label{GUTPlanckpos}
\eeq
This is the positive-$k$ analogue of the relation (\ref{GUTPlanck})
that we previously found for negative $k$.

Note that by itself,
the relation (\ref{gutresult}) generally requires
$M > M_{\rm GUT}$.  In other words,
although our gauge coupling unification mechanism successfully
produces the expected values of 
gauge couplings at the fundamental mass scale $M$,
this scale $M$ must {\it exceed}\/ the usual GUT scale.
It is easy to understand physically why this is the
case.  Looking back at (\ref{powerlaw}) and replacing $\hat k\to -k$
(with $k$ a positive quantity), we see that the sign of the anomaly-induced
contributions is flipped relative to the negative-$k$ case.
However, this is satisfactory 
as an explanation of the gauge couplings at
a scale $M$ when $M > M_{\rm GUT}$;
indeed, if we imagine running the three gauge couplings
 {\it past}\/ their usual unification point, they begin
to split again in opposite directions.
Of course, the value of $M$ is itself an unphysical quantity, since
it can always be rescaled by introducing an additional
overall rescaling factor into the 
five-dimensional bulk metric.  As discussed in Sect.~2, such
an overall rescaling does not change the physics.  There
is therefore nothing improper about having $M>M_{\rm GUT}$. 
To see this more explicitly, let us imagine
introducing an additional overall rescaling factor $e^{\sigma_0}$
into the five-dimensional metric.
We would then find that $M$ is replaced by 
$e^{\sigma_0/2} M$ in the above expressions,
and, for suitable choices of $\sigma_0$ 
we can always bring $M$ below $M_{\rm GUT}$.
Thus, our mechanism for gauge coupling unification continues to work,
even with $M<M_{\rm GUT}$.
However, just as in the negative-$k$ solution, there remains
a difficulty making this solution consistent with electroweak
symmetry breaking.  This issue will be discussed further in
Sect.~4.


\section{Electroweak symmetry breaking:  An open question}
\setcounter{footnote}{0}

In the previous section, we discovered that gravity-localized
compactifications automatically predict the correct values
of the low-energy gauge couplings
as a result of a rescaling anomaly. A natural question that remains 
to be answered is the origin of the electroweak scale.  Let us 
consider the Higgs potential
\beq
         V(H) ~=~ -m_0^2 |H|^2 +\lambda |H|^4
\label{Higgspot}
\eeq
where the minimum of the potential is located at the value 
$v_0\equiv m_0/\sqrt{2\lambda}$. 
In the positive-$k$ scenario, 
the mass parameter $m_0$ (which is assumed to be near the Planck scale) 
is naturally red-shifted towards the TeV scale.  Thus, as proposed in Ref.~\cite{RS1},
the physical electroweak symmetry breaking
scale $v\approx 246$ GeV is identified as $v\equiv e^{-\pi k R/2} v_0$.
In this way one obtains a natural electroweak 
symmetry breaking scale~\cite{RS1} starting from a higher fundamental scale.

However, in the negative-$k$ scenario, 
we see that if the Higgs mass parameter
$m_0$ is near the TeV-scale, then the corresponding classical blue-shift 
rescales the Higgs mass parameter to be near the Planck scale.
Thus, in order to obtain
the correct Higgs mass parameter at the TeV-scale, we would need to begin
with an initial mass parameter 
$m_0\approx 10^{-4}$ eV!
Curiously, however, this is believed to be the size of 
the four-dimensional cosmological constant.
Thus, we are led to the rather unorthodox idea that perhaps
the non-zero cosmological constant triggers electroweak symmetry-breaking.
Of course, in such gravity-localized scenarios the cosmological-constant
problem is intimately connected with the problem of radius stabilization.
Discussions of these issues can be found in Ref.~\cite{Kaloper}.

A second possible way of avoiding these undesirable blue-shifting
effects might be to 
eliminate the ``bare'' Higgs mass parameter altogether, essentially 
setting $m_0=0$. 
Of course, we would still require a mechanism for
triggering electroweak symmetry breaking, but here one could
conceivably use the Coleman-Weinberg mechanism~\cite{CW} wherein
the required Higgs potential is generated via radiative corrections.
This leads to the Coleman-Weinberg effective potential
\beq
     V(H)+\delta V_{\rm anomaly}  ~=~ \lambda \,|H|^4 ~+~ 
         C \,|H|^4 \, \ln \left( |H|^2/\mu^2\right) -2 C \Lambda |H|^4~
\label{CWpotential}
\eeq
where $\mu$ is a renormalization scale, $C$ is a model-dependent
constant, and $\lambda$ is now an effective
coupling whose value depends on the renormalization scale $\mu$.  
Note that the last term in the potential (\ref{CWpotential}) is the
anomalous contribution that arises due to the rescaling of the Higgs fields.
As we saw in Sect.~3 for the gauge couplings, this anomaly can viewed as 
effectively rescaling the renormalization scale 
$\mu\rightarrow \mu e^{\Lambda}$. 
Thus, although the electroweak symmetry continues to be broken
by radiative corrections, the coupling $\lambda$ is now defined
at the renormalized scale $\lambda=\lambda(\mu e^{-\pi {\hat k} R/2})$. 
However, as discussed in Sect.~3, defining the coupling at a rescaled 
renormalization point is equivalent to shifting the coupling by a finite 
amount. In particular, the minimum for a simple $\lambda/4!\phi^4$ theory 
with $C=\lambda^2/(256\pi^2)$~\cite{CW} becomes
\beq
    \langle H \rangle ~\approx~  \mu \,
       e^{-16\pi^2/(3 \lambda(\mu e^\Lambda))} ~=~ 
        v_0 \,e^{\pi {\hat k} R/2}  
\eeq
where $v_0$ is the minimum that would have arisen without the anomaly 
contribution. Thus, we see that even in the Coleman-Weinberg scenario,
the electroweak symmetry breaking scale $v\equiv \langle H \rangle$ is 
again rescaled. Indeed, this is the quantum analogue of the
classical rescaling of $m_0$.  
Of course, at higher loops, the blue-shifting may also receive 
contributions from the anomalous dimensions of quantum fields. 
Furthermore, just as in the usual Coleman-Weinberg 
scenario, the presence of a heavy top quark mass continues to effectively 
destabilize the potential.  We therefore leave this issue for further study.

Thus, given this rescaling of $v$, we see 
that we now have three 
simultaneous equations that relate the three physical observables 
$M_{\rm Planck}$, $M_{\rm GUT}$, and $v$ to the three 
parameters ($M$, $k$, and $kR$) that define our gravity-localized
compactification.
Assuming $kR\gsim {\cal O}(10)$, so that we may approximate
$2\sinh(\pi k R)\approx \exp(\pi k R)$,
we find that these three simultaneous equations take the form
\beqn
         k>0:&&~~~~~~~~~\cases{
                M_{\rm Planck} = M \sqrt{M/k}\, \exp(\pi k R/2) & \cr
                M_{\rm GUT} = M  \exp(-\pi k R/2) & \cr
                v = M  \exp(-\pi k R/2) & \cr} \nonumber\\
         k<0:&&~~~~~~~~~\cases{
                M_{\rm Planck} = M \sqrt{M/\hat k}\, \exp(\pi \hat k R/2) & \cr
                M_{\rm GUT} = M  \exp(\pi \hat k R/2) & \cr
                v = M  \exp(\pi \hat k R/2)~. & \cr} 
\label{fundeqs}
\eeqn
In principle, it is therefore possible to solve simultaneously in each case.  
For example, 
solving the Planck and GUT equations
simultaneously in each case yields the 
relations (\ref{GUTPlanckpos}) and (\ref{GUTPlanck}).    
Likewise, in the positive-$k$ solution, if we take $M\approx k$ for simplicity
and simultaneously solve the 
Planck and electroweak constraints, we obtain
the solution $kR\approx 12$, $M\approx 10^{10}$ GeV.

Unfortunately, we now see that the additional GUT constraint is incompatible
with both of these conclusions.
Indeed, from (\ref{fundeqs}), we see that regardless of the sign of $k$,
these constraint equations together imply that 
\beq
                          M_{\rm GUT} ~\approx~ v~.
\eeq
Clearly, this is patently false, failing by approximately 14 orders of magnitude.
We stress again that this conclusion follows directly from the wavefunction
rescalings and their associated anomalies, which in turn follow directly from the
very nature of the gravity-localized compactifications.
While this conclusion might be altered if we are willing to accept a large
a hierarchy between $v_0$ and $M$,
this new hierarchy would have to be 
exactly as large as the hierarchy between $M_{\rm GUT}$ and $v$ that 
we are seeking to explain.
Thus, we conclude that we cannot simultaneously
generate the Planck/electroweak hierarchy {\it and}\/ explain gauge 
coupling unification in such gravity-localized compactifications.
Indeed, this result holds regardless of the sign of $k$.
This, then, seems to be a major difficulty of the gravity-localization framework.

To some extent, this state of affairs is not surprising.
In the $k>0$ scenario, as advocated in Ref.~\cite{RS1},
mass scales on the Standard-Model brane are red-shifted
down from a high fundamental scale in the bulk.
Thanks to the contribution from the rescaling anomaly,
this includes not only the classical (bare) mass scales
that appear directly in the Lagrangian of the Standard-Model D3-brane,
but also those ``quantum'' scales (such as the GUT scale) which
are generated by quantum effects.
Therefore, while this scenario provides an elegant solution
to the hierarchy problem by generating the electroweak
scale from the Planck scale, this scenario is generally incapable
of explaining physics that requires the presence of high scales such
as the GUT scale.
Our arguments concerning gauge coupling unification in this
scenario make this last point particularly explicit.
By contrast, the $k<0$ scenario has a different complexion.
Here all mass scales on the Standard-Model D3-brane
are subjected to a {\it blue-shifting}\/ effect
which, when coupled with the effects of the rescaling anomaly,
simultaneously {\it raises}\/ the classical and quantum
mass scales on the Standard-Model D3-brane.
As we have seen, this scenario thus has no trouble
accommodating gauge coupling uniifcation, which relies
upon having a high GUT scale.  Indeed, 
as we shall shortly see, the entire GUT structure  
of the Standard Model (such
as the correct neutrino masses and proton decay) survives intact.
However, this scenario has trouble explaining those features
that ordinarily rely on the presence of a low electroweak scale
(such as the mass scales appearing in Higgs potential).
 
Thus, these two scenarios are in some sense complementary,
with neither providing a full and simultaneous explanation of
the disparate energy scales in the Standard Model.
In fact, it would appear that this shortcoming is a generic
feature of such gravity-localized compactifications.
By their very nature, the effect of gravity localization
is to introduce a warp rescaling factor (of whatever sign)
into the metric on the Standard-Model brane.
However, such a warp rescaling factor is universal,
and will affect all Standard-Model mass scales simultaneously.  Thus,
such warp factors do not seem to have the flexibility that
would be required in order to simultaneously explain
physics that relies on the existence of two disparate
mass scales on the same brane.

\section{Brane/bulk couplings and neutrino masses}
\setcounter{footnote}{0}

In this section, we investigate the issue of brane/bulk couplings in 
gravity-localized compactifications.  In traditional extra-dimension 
compactifications with product spacetimes, such couplings between brane fields
and bulk fields have proven to play a crucial role in explaining
the possible origin of small numbers such as neutrino 
masses~\cite{DDGneutrinos,ADDneutrinos}.
Generally,
these small numbers emerge thanks to 
a suppression factor arising as a result of the large volume of the extra dimension.
It is therefore important to understand how the sizes of
these brane/bulk couplings
are modified when gravity is localized and no large radii for the
extra dimensions are required.

Although the following considerations are quite general and
apply to a variety of situations, for concreteness we shall   
restrict our attention to the case that is most relevant for
neutrino masses.
In Refs.~\cite{DDGneutrinos,ADDneutrinos}, it was shown that
small phenomenologically viable neutrino masses could
be produced in extra-dimension scenarios 
by considering the right-handed neutrino
(a Standard-Model singlet field) to reside in the
bulk rather than on the brane containing the Standard Model.
To this end, we shall consider the particular 
case of a coupling between
two brane
fields (a left-handed neutrino $\nu_L$ and a Higgs field $H$)
and a single bulk field (a 
``right-handed'' neutrino field $\Psi\equiv (\psi_1,{\bar \psi_2})^T$ 
in the Weyl basis).
As in Ref.~\cite{DDGneutrinos}, we shall take $\Psi$ to satisfy 
the orbifold relations
$(\psi_1,\psi_2)\to (\psi_1,-\psi_2)$ under $y\to -y$,
as a result of which only $\psi_1$ can couple to the
left-handed neutrino $\nu_L$ located on the brane at the
orbifold fixed point $y=\pi R$.  
The relevant terms in the action for this system are then
given by
\beqn
      S  &=& \int d^4 x \,\sqrt{-g} \, \biggl\lbrace
       g^{\mu\nu} D_\mu H^\dagger D_\nu H
     ~+~ e^\mu_a \bar\nu_L i \bar\sigma^a D_\mu \nu_L ~+~
     y_\nu (H \nu_L \psi_1|_{y=\pi R} + {\rm h.c.})\biggr\rbrace \nonumber\\
       &&~~~+ \int d^4 x \int_{-\pi R}^{\pi R} dy\,
        \sqrt{-G} \left( M e^M_A  \bar\Psi i \gamma^A 
        \partial_M \Psi \right)~.
\label{neutrinoaction}
\eeqn
In the first line we have given the kinetic-energy terms for the
Higgs field and the left-handed neutrino, where $g\equiv g^{(\pi R)}$ is
the metric on the Standard-Model D3-brane located at $y=\pi R$
and where $e^\mu_a$ is the vierbein necessary in order to compensate
for the non-flat metric.
We have also given the Dirac coupling between the left-handed 
neutrino, the Higgs field, and the right-handed neutrino, where $y_\nu$
is the Yukawa coupling.
We shall generally assume ${\cal O}(10^{-6}) ~\lsim~ y_\nu~\lsim~{\cal O}(1)~$
in order to reflect the expectation that the neutrino Yukawa couplings
are within the ranges already set by their corresponding $SU(2)$ lepton 
counterparts, with $y^{(e)}_\nu\approx {\cal O}(10^{-6})$ for the electron neutrino
and $y^{(\tau)}_\nu \approx {\cal O}(1)$ for the tau neutrino.
Of course, these values are only meant to serve as approximate guides.
Finally, in the second line of (\ref{neutrinoaction})
we have given the bulk kinetic-energy term for
the right-handed fermion $\Psi$, where $G$ is the full bulk five-dimensional
metric and $M$ is the overall fundamental 
mass scale in the bulk. 

\subsection{The $k>0$ scenario}

In order to see the emergence of a ``volume'' suppression for the
Dirac brane/brane/bulk coupling, the next step is to rescale the fields 
in the system so that they all
have canonically normalized kinetic-energy terms.
Let us first do this for the case of the scenario proposed in
Ref.~\cite{RS1}, where the full metric is given in (\ref{trial}) 
with the solution (\ref{DDGsoln}).
Recalling that $e^\mu_a$ and $e^M_A$ scale like $\sqrt{g^{\mu\nu}}$ and
$\sqrt{G^{MN}}$ respectively, we find that 
the effective four-dimensional action describing our system 
takes the form
\beqn
   S &=&  \int d^4 x \, \biggl\lbrace
       e^{- \pi k R}  D_\mu H^\dagger D^\mu H
     ~+~ e^{-3 \pi k R/2} \bar\nu_L i \bar\sigma^\mu D_\mu \nu_L 
     ~+~ e^{-2 \pi k R} y_\nu (H \nu_L \psi_1^{(0)}
        + {\rm h.c.}) \nonumber\\     
            && ~~~~~~~~
     ~+~ {4 M\over 3k} \sinh\left( {3\pi k R\over 2}\right)  \bar \psi_1^{(0)} 
        i\bar\sigma^\mu \partial_\mu \psi_1^{(0)}\biggr\rbrace~.
\label{action2}
\eeqn 
In (\ref{action2}), we have kept only the zero-mode $\psi_1^{(0)}$ for 
simplicity,
as this is sufficient for deducing the effect of the volume factor.
We have also integrated over the fifth dimension in order to 
derive the last term, and we shall henceforth approximate 
$2\sinh(3\pi k R/2) \approx \exp(3\pi k R/2)$.
Thus, in order to canonically normalize the kinetic-energy terms, we must
do a Weyl-rescaling of the wavefunctions.
This causes our brane/bulk coupling term to take the form
\beq
         \int d^4 x ~ e^{-3 \pi k R/2} \sqrt{ 3k\over 2M} \,y_\nu 
        (H \nu_L \psi_1^{(0)}+{\rm h.c.})~,
\eeq
whereupon we see that the effective Dirac neutrino mass $m_\nu$ is given by
\beq
         m_\nu ~\approx~ 
        \sqrt{3k\over 2M} ~y_\nu \langle H \rangle\, e^{-3\pi kR/2}~.
\eeq
We thus obtain essentially the expected result, 
multiplied by an extra ``warp'' suppression factor $e^{-3\pi kR/2}$. 
Unfortunately, the effects of this warp factor are quite severe
in the scenario of Ref.~\cite{RS1}:  taking $kR\approx 12$
(as appropriate for the solution to the gauge hierarchy problem
in Ref.~\cite{RS1}), we find $ e^{-3\pi kR/2}\approx 10^{-24}$.
Thus, with $\langle H\rangle \approx {\cal O}(10^2)$ GeV and $y_\nu\approx 
 {\cal O}(10^{-6})$,
we find 
\beq
         m_\nu ~\approx ~ 10^{-24} \,y_\nu \langle H\rangle ~\approx ~
          10^{-19}~{\rm eV}~,
\eeq
which is far too small to be the correct neutrino mass.
Even if we take the extreme case $y_\nu\approx {\cal O}(1)$, 
we find $m_\nu\approx 10^{-13}$ eV, which is still too small to match
current experimental expectations.
Of course, for a proper treatment one must include the effects of
 {\it all}\/ of the Kaluza-Klein modes of the bulk $\Psi$ field, as 
in Ref.~\cite{DDGneutrinos}, and diagonalize an infinite-dimensional 
mass matrix.  In the case of the gravity-localized scenario, such 
a Kaluza-Klein decomposition would presumably follow 
along the lines of Refs.~\cite{RS2,GWbulkfields};  however,
this is beyond the scope of the present paper.  
Nevertheless, just at the level of the Dirac mass coupling,
it is already apparent 
that in the scenario of Ref.~\cite{RS1},
the warp factor from the metric generically tends 
to {\it over-suppress}\/ the brane/bulk couplings.
This is therefore an important phenomenological
problem for this scenario.

\subsection{The $k<0$ scenario}

Let us now see how this result is altered in our modified scenario with $k<0$.
We begin again with the action (\ref{neutrinoaction}), and now substitute the bulk metric given in (\ref{DDGmetricnegative}).
Following the same steps as before, we find that
this leads to the effective four-dimensional
action obtained by replacing $k\to -\hat k$ in (\ref{action2}).
Because this flips the warp factors in the first line of (\ref{action2})
while leaving the second line invariant,
the required Weyl-rescaling of the brane-field wavefunctions 
is flipped while the Weyl-rescaling of the bulk-field wavefunction is unaltered. 
We thus find that our effective Dirac neutrino mass $m_\nu$ is now given by
\beq
         m_\nu ~\approx~ 
        \sqrt{3\hat k \over 2 M} ~y_\nu \langle H \rangle~
\eeq
where $M_{\rm Planck}$ is given by (\ref{MPlanckDDG}).
Of course, we have already seen in (\ref{GUTPlanck}) that
the ratio $\hat k/M\approx 10^{-4}$ is fixed by the ratio of the GUT scale
to the Planck scale.
Taking $y_\nu\approx {\cal O}(10^{-6})$ as a rough estimate for the 
electron-neutrino Yukawa coupling, we thus find 
\beq
       m_\nu ~\approx~ 10^{-2}\, y_\nu \langle H\rangle
     ~\approx~ 10^{3} ~{\rm eV}~.
\eeq
This too fails to be within the range of the current experimental expectations.
Thus, we see that while the scenario of Ref.~\cite{RS1} yields Dirac neutrino
masses that are vanishingly small,
in our scenario the Dirac neutrino masses are slightly larger than expected.

Of course, as in any model of neutrino masses, 
the ultimate values of the neutrino masses
depend crucially on the chosen values of the corresponding Yukawa couplings. 
For example, in order to obtain a value $m_\nu \approx 10^{-3}$ eV, we would need a 
Dirac Yukawa coupling $y_\nu \approx 10^{-12}$.   
Alternatively, one can relax
the condition that the ultraviolet cutoff scale on the Standard-Model D3-brane
be the same as the physical scale $M$ in the bulk. 
Indeed, as discussed in Ref.~\cite{Verlinde}, one can imagine attempting
to realize such gravity-localized scenarios within the framework
of Type~I string theory.
In such cases, it is reasonable to assume that our brane ultraviolet cutoff 
is of the order of the string scale $M_I$, which in turn
can be substantially different from the five-dimensional bulk Planck mass 
$M$.  For example, taking $M=10$ TeV and an ultraviolet cutoff scale $M_I=10^{10}$
GeV, we can obtain a neutrino mass of $10^{-3}$ eV for a Yukawa coupling
$y_\nu \approx 10^{-6}$.  The usual Planck and GUT scales continue to be 
related provided $\sqrt{M/\hat k}\approx 10^{8}$ and 
$\hat k R \approx 10$, which gives rise to the solutions 
$\hat k\approx 10^{-3}$ eV
and $R^{-1} \approx 10^{-4}$ eV. 
Remarkably, we thus obtain a millimeter-sized extra dimension as in Ref.~\cite{ADD}!
Obviously, there are now large hierarchies between the different mass 
scales in the theory,
but this gives an example of how the neutrino problem might be solved.
Thus,
while a problem exists, it seems that our scenario with $k<0$ 
can perhaps be more easily accommodated within a flavour-violating theory
of neutrino masses.

We have seen that obtaining Dirac neutrino masses within the 
required experimental range does not work well for both scenarios. However,
since the blue-shifting allows us to bring high-scale mechanisms down to
lower energies (such as occurred for gauge coupling unification),
it follows that the {\it usual}\/ 
seesaw mechanism can also be brought down to low energies in a similar way.

To see this,
let us consider the usual seesaw mechanism on the Standard-Model brane.
Unlike the previous discussion, we shall here introduce
a right-handed neutrino {\it only}\/ on the brane. The
wavefunction of the 
right-handed neutrino scales just like that for the left-handed neutrino,
and after rescaling, the Lagrangian takes the form
\beq
         \int d^4 x ~ \,y_\nu H \nu_L \psi_1
             + M e^{\pi {\hat k} R/2} \psi_1\psi_1 +{\rm h.c.}  ~.
\eeq
Here $\psi_1$ is now a purely four-dimensional Weyl spinor on the
Standard-Model brane, and $M$ is a Majorana mass for the right-handed
neutrino field.  Note that only the bare Majorana mass term is rescaled
since this term classically breaks the scale invariance. Integrating out
the right handed neutrino field then gives rise to a dimension-five operator 
\beq
     {y_\nu^2 \over M e^{\pi {\hat k} R/2}} \,\nu_L \nu_L H H 
\eeq
which in turn leads to a neutrino Majorana mass 
\beq
    m_\nu ~\approx~ {y_\nu^2 \langle H\rangle^2  \over 
        M e^{\pi {\hat k} R/2}} ~\approx~ 10^{-3} ~{\rm eV}~, 
\eeq
where we have used $M=10$ TeV, ${\hat k} R \approx 18$, and 
$y_\nu \approx 1$.  Thus, for the negative-$k$ scenario,
we see that we are able to obtain the 
required neutrino masses using the usual seesaw mechanism
without ever having to introduce a heavy mass scale. 
This arises because the negative-$k$ scenario, due to its blue-shifting
factor, essentially reproduces the usual GUT structure and mass relations
on the Standard-Model brane even though the 
bare mass scales on the Standard-Model brane are small.
Indeed, similar arguments
can also be used to demonstrate that proton stability is also not 
a problem in the negative-$k$ scenario. 

Note that in the positive-$k$ scenario, the simple 
seesaw mechanism does not work 
since the large intermediate mass scale will be red-shifted rather
than blue-shifted. This leads
to unaccepatble levels of lepton-number violation at low energies.
A similar problem will also arise for proton decay.
Thus, we see that the negative-$k$ scenario can more easily 
accommodate neutrino masses within the required experimental range.

\section{Conclusions}
\setcounter{footnote}{0}

In this paper, we have considered gravity-localized 
compactification scenarios in which a ``warp factor'' generates 
the hierarchy between the weak scale and the usual four-dimensional 
Planck scale. 
Like the scenario originally proposed in Ref.~\cite{RS1},  
 only one extra dimension is required,
and there are no large hierarchies between the scale 
       of extra dimensions and the fundamental physical scale of the theory.
Unlike the scenario of Ref.~\cite{RS1},
however, in our formulation this scenario involves no high physical scales 
(either on the brane or in the bulk).

Our main result is 
that gauge coupling unification emerges naturally in such
scenarios, and arises thanks to a rescaling anomaly.
This ``anomaly-induced'' gauge coupling unification thus 
explains the different values of the low-energy gauge couplings
on the Standard-Model brane,
and represents
a new mechanism for achieving gauge ``unification'' at reduced energy
scales.  Because of its generality, requiring only the presence
of a non-trivial warp factor,
this anomaly-induced unification mechanism
may also be applicable to many other similar      
gravity-localized scenarios that have recently been 
proposed~\cite{others}.

Given this result, we then proceeded to investigate
the compatibility of the GUT scale and the electroweak
symmetry breaking scale.  Unfortunately, the results are rather
discouraging, generically requiring $M_{\rm GUT}\approx v$.
This is a signal that gravity-localized compactification
scenarios generically cannot accommodate widely separated
mass scales on a single Standard-Model brane.
(By contrast, the Planck scale emerges directly from the
bulk where gravity is free to propagate, even if in
a restricted manner.)
We also pointed out various speculative ideas 
concerning how this issue might ultimately be resolved, and
also considered the sizes of generic brane/bulk couplings in
such gravity-localized scenarios.

Overall, despite the difficulties in accommodating the
GUT scale and the electroweak scale simultaneously,
we feel that the phenomenological prospects of gravity-localization
are rich and have hardly been explored.
The multitude of possibilities
inherent in this framework therefore suggests 
that this question is worthy of further study.

\bigskip
\medskip
\leftline{\large\bf Acknowledgments}
\medskip

ED wishes to thank C.~Grojean and J.~Mourad
for useful discussions. 
KRD and TG also wish to acknowledge the hospitality of the
Aspen Center for Physics where part of this work was done.


\setcounter{section}{0}

\vfill\eject
\bigskip
\medskip

\bibliographystyle{unsrt}

\end{document}